\documentclass[aps,twocolumn]{revtex4-1}
\usepackage{bm}
\usepackage{amsmath}
\usepackage{amssymb} 
\usepackage{epsfig}
\usepackage{graphicx}
\usepackage{color}
\usepackage{xcolor}
\usepackage[normalem]{ulem}
\usepackage{cancel}
\usepackage[unicode=true,colorlinks=true,citecolor=blue]{hyperref}
\usepackage{eufrak}
\newcommand{\pderiv}[2]{\frac{\partial #1}{\partial #2}}
\renewcommand{\phi}{\varphi}
\renewcommand{\kappa}{\varkappa}

\renewcommand{\i}{\mathrm i}
\newcommand{\e}{\mathrm e}
\newcommand{\eps}{\varepsilon}

\newcommand{\aver}[1]{\left \langle #1 \right \rangle}

\begin{document}

\title{
Faraday and Kerr rotation due to photoinduced orbital magnetization in two-dimensional electron gas}

\author{M.\,V.\,Durnev}

\affiliation{Ioffe Institute, 194021 St.\,Petersburg, Russia}

\begin{abstract}

We study theoretically the Faraday and Kerr rotation of a probe field due to the orbital magnetization of a two-dimensional electron gas induced by a circularly polarized pump. We develop a microscopic theory of these effects in the intraband spectral range based on the analytical solution of the kinetic equation for linear and parabolic energy dispersion of electrons and arbitrary scattering potential. We show that the spectral dependence of rotation angles and accompanying ellipticities experiences a sharp resonance when the probe and pump frequencies are close to each other. At the resonance, the Faraday and Kerr rotation angles are of the order of $0.1^\circ$ per 1~kW/cm$^2$ of the pump intensity in graphene samples, corresponding to a pump-induced synthetic magnetic field of about 0.1~T. We also analyze the influence of the dielectric contrast between dielectric media surrounding the two-dimensional electron gas on the rotation angles.
\end{abstract}

\maketitle

\section{Introduction}

Optically induced magnetization and its manipulation in solids have recently attracted significant attention in solid-state physics~\cite{Kirilyuk2010, Stupakiewicz2017, Chai2017, Cheng2020}. Absorption of circularly polarized photons results in efficient magnetization of electron and hole systems in the process of optical spin orientation through both the interband and intraband optical transitions~\cite{spin_phys_book, Kusraev2008, Glazov2012, Ganichev2002, Ivchenko2004, Murdin2006}. Besides the spin orientation, the circularly polarized light induces orbital currents of charge carriers, and hence, the orbital magnetic moment, known as the inverse Faraday effect (IFE)~\cite{Pitaevskii1961, Ziel1965}. The orbital magnetization due to the IFE is being actively studied in different systems, including metals and semiconductors~\cite{Hertel2006, Battiato2014, Berritta2016, Ryzhov2016}, ferromagnets~\cite{Kimel2005}, superconductors~\cite{Mironov2021}, metallic nanoparticles~\cite{ Hurst2018} and graphene~\cite{Tokman2020}.

To probe the light-induced orbital magnetic moment, one can use the pump-probe Faraday and Kerr spectroscopy -- the method, which is widely employed to study the magnitude and dynamics of magnetization related to both spin and orbital magnetic moment~\cite{Awschalom1985, Zheludev1994, Kato2004, Crooker2005, Greilich2006, Glazov2010, Glazov2012, Passmann2018, Cheng2020}. In this method, one measures the rotation of the polarization plane of linearly polarized probe beam, which is reflected from or transmitted through the medium with pump-induced magnetization. While the theory of the pump-probe Faraday and Kerr effects due to spin magnetization has been developed for bulk and low-dimensional semiconductor systems~\cite{Aronov1973, Svirko1994, Yugova2009, Glazov2012}, consistent microscopic theory of these effects due to orbital magnetization is still missing. The naive mechanism of such a Faraday rotation could involve magnetic field induced by the orbital currents, however this magnetic field is extremely small and, hence, cannot be the major source of rotation. The third-order contribution to ac current induced by elliptically polarized electric field in graphene and responsible for the Faraday rotation, has been calculated in Ref.~[\onlinecite{Glazov2014}]. However, the calculations were based on a simplified relaxation model, which does not fully capture the specifics of electron scattering in two-dimensional systems.

\begin{figure}[htpb]
\includegraphics[width=0.3\textwidth]{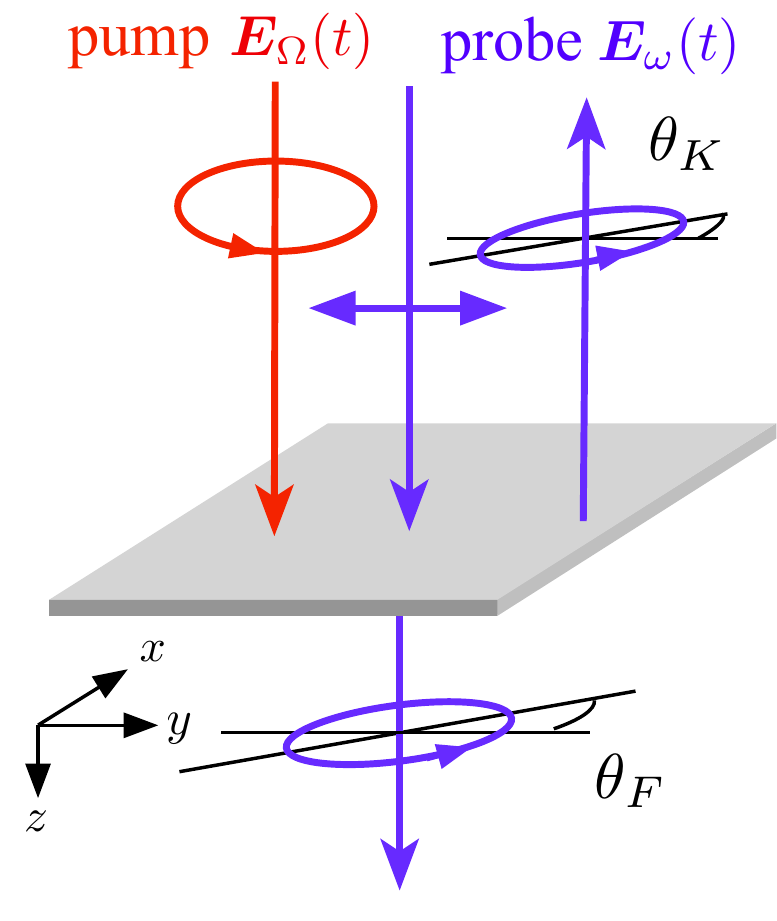}
\caption{\label{fig1} Schematic picture of the pump-induced Faraday and Kerr rotation in the two-dimensional electron gas. Electric field of the circularly polarized pump acts as a synthetic magnetic field resulting in the rotation of the linearly polarized probe field. $\theta_F$ and $\theta_K$ are the Faraday and Kerr rotation angles, respectively.
 }
\end{figure}

Here, we study the Faraday and Kerr rotation due to the orbital magnetization induced by circularly polarized pump in a two-dimensional electron gas (2DEG). We show that the circularly polarized electric field of the pump modifies the high-frequency conductivity of 2DEG, resulting in the circular birefringence and dichroism. This, in turn, leads to rotation of the transmitted and reflected probe field. Moreover, the initially linearly polarized probe becomes elliptically polarized (acquires ellipticity), Fig.~\ref{fig1}. 
We develop a microscopic theory of the pump-induced high-frequency conductivity of 2DEG due to intraband optical transitions and calculate the Faraday and Kerr angles as well as the corresponding ellipticities. 
The theory accounts for electron scattering by impurities and describes both non-absorbing and absorbing regimes of the pump and probe fields. We derive analytical expressions for the Faraday and Kerr angles and ellipticities valid for parabolic and linear energy dispersion of 2D electrons and arbitrary scattering potential. We also analyze the influence of the dielectric contrast between dielectric media surrounding 2DEG on the rotation angles.

We show that the spectral dependence of rotation angles and ellipticities experiences a sharp resonance, when probe and pump frequencies are close to each other. The width and the magnitude of resonance are determined by a long energy relaxation time, rather than a short momentum relaxation time.
At the resonance, and at $\Omega \tau_1 \sim 1$, where $\Omega$ is the pump frequency, and $\tau_1$ is the momentum relaxation time, the Faraday and Kerr rotation angles are of the order of $0.1^\circ$ per 1~kW/cm$^2$ of the pump intensity in graphene samples. We also calculate a synthetic magnetic field, an effective magnetic field, which leads to the same rotation angles as the circularly polarized pump. In graphene samples, this synthetic magnetic field amounts to $\sim 0.1$~T per 1~kW/cm$^2$ of the pump intensity at $\Omega \tau_1 \sim 1$. 

\section{Faraday and Kerr rotation by a 2D conducting medium}

We consider 2DEG occupying  the plane $z=0$ and surrounded by dielectrics with refractive indices $n_1$ at $z <0$ and $n_2$ at $z > 0$. The 2DEG is irradiated by normally incident pump and probe beams with electric fields $\bm E_\Omega(t) = \bm E_\Omega \e^{ - \i \Omega t} + \mathrm{c.c}$ and $\bm E_\omega(t) = \bm E_\omega \e^{- \i \omega t} + \mathrm{c.c}$, respectively, see Fig.~\ref{fig1}.  
In the absence of pump field, $E_\Omega = 0$, the probe field induces electric current in 2DEG $\bm j(t) = \bm j_\omega \e^{- \i \omega t} + \mathrm{c.c}$, which oscillates at the probe frequency and is parallel to the probe electric field $\bm E_\omega$. The current is related to the probe field as $\bm j_\omega = \sigma \bm E_\omega$, 
where $\sigma = e^2 n_e \tau_1/m(1-\i\omega\tau_1)$ is the high-frequency 2DEG conductivity, $e$ and $m$ are the electron charge and effective mass, respectively, $n_e$ is the 2D electron concentration and $\tau_1$ is the momentum relaxation time. 

In the presence of the pump field, the third-order contributions to the current $\bm j_\omega$ appear. These contributions in the isotropic 2DEG are described by the following equation with three complex parameters $\gamma_j$~\cite{Durnev2021}:
\begin{multline}
\label{phen}
\bm j_\omega = \gamma_1 |\bm E_\Omega|^2 \bm E_\omega + \gamma_2 \left[ \bm E_\Omega^* (\bm E_\Omega \cdot \bm E_\omega) + \bm E_\Omega (\bm E_\Omega ^*\cdot \bm E_\omega) \right] \\
+ \i \gamma_3 \left[ \bm E_\omega \times \left[ \bm E_\Omega \times \bm E_\Omega^* \right] \right]\:.
\end{multline}
Here, $\gamma_1$ describes the change of isotropic conductivity due to the pump radiation, whereas $\gamma_2$ and $\gamma_3$ give rise to the transverse current in the direction perpendicular to $\bm E_\omega$ induced by linearly and circularly polarized pump, respectively. In this paper, we consider circularly polarized pump, and therefore, the $\gamma_3$ contribution~\footnote{Linearly polarized pump causes linear birefringence and dichroism of 2DEG, which manifests itself in different transmission, reflection and absorption of the probe field polarized parallel and perpendicular to the pump field.}. For circularly polarized pump, Eq.~\eqref{phen} yields the transverse current described by the off-diagonal conductivity  $\sigma_{xy} = -\sigma_{yx} = \gamma_3 |\bm E_\Omega|^2 P_{\rm circ}$, where $P_{\rm circ} = \pm 1$ for right-hand and left-hand circular polarization, respectively. Note that, when the probe field is static, i.e. at $\omega = 0$, the $\gamma_3$ contribution describes the appearance of a transverse direct current in the presence of a circularly polarized pump -- the so-called photovoltaic or circular Hall effect~\cite{McIver2020, Durnev2021, Candussio2022}.

Pump-induced transverse conductivity $\sigma_{xy} = -\sigma_{yx}$ leads to circular birefringence and circular dichroism, i.e. different transmission and absorption of the right-hand and left-hand circularly polarized components of the probe field. The incident linearly polarized probe field is a superposition of circularly polarized fields $\bm E_{\omega,\pm}^{(i)} = E_\omega^{(i)} \bm o_\pm$, where $\bm o_\pm$ are circularly polarized unit vectors related to the unit vectors $\bm e_x \parallel x$ and $\bm e_y \parallel y$ as $\bm o_\pm = (\bm e_x \pm \i \bm e_y)/\sqrt{2}$. The amplitude transmission and reflection coefficients of $\bm E_{\omega,\pm}^{(i)}$ are given by~\cite{Chiu1976} 
\begin{equation}
\label{rt}
t_\pm = \frac{t_{12}}{1 + \alpha_\pm}\:,~~~r_\pm = \frac{r_{12} - \alpha_\pm}{1 + \alpha_\pm}\:,
\end{equation}
where $r_{12} = (n_1-n_2)/(n_1+n_2)$ and $t_{12} = r_{12} + 1$ are the amplitude reflection and transmission coefficients for the light incident on the boundary between two dielectrics in the absence of the 2DEG layer, $\alpha_\pm = 2\pi\sigma_\pm/(c\bar{n})$, $\sigma_\pm = \sigma_{xx} \pm \i \sigma_{xy}$, $\bar{n} = (n_1+n_2)/2$, and $c$ is the speed of light in vacuum.  

Pump-induced anisotropy of the transmission and reflection coefficients leads to the rotation of the linear polarization of the transmitted and reflected probe fields. We will further consider the low-intensity regime, when the pump-induced off-diagonal conductivity is much smaller then the diagonal one, i.e. $|\sigma_{xy}| \ll |\sigma_{xx}|$, and $\sigma_{xx} \approx \sigma$. In that case the differences $t_+ - t_-$ and $r_+ - r_+$ are much smaller than the corresponding sums, and the Faraday rotation angle $\theta_F$ and the ellipticity $\epsilon_F$ of the transmitted probe field are~\cite{Palik1970, OConnell1982, Glazov2012} 
\begin{equation}
\label{thetaF}
\epsilon_F - \i \theta_F \approx \frac{t_+ - t_-}{t_+ + t_-}\:.
\end{equation} 
Analogously, the Kerr rotation angle $\theta_K$ and the accompanying ellipticity $\epsilon_K$ of the reflected probe field are given by
\begin{equation}
\label{thetaK}
\epsilon_K - \i \theta_K \approx  \frac{r_+ - r_-}{r_+ + r_-} \:.
\end{equation} 

By substituting Eq.~\eqref{rt} to Eqs.~\eqref{thetaF} and \eqref{thetaK}, we obtain
\begin{equation}
\label{thetaF2}
\theta_F + \i \epsilon_F \approx  \frac{2\pi \sigma_{xy}}{c \bar{n}(1 + \alpha)}\:,
\end{equation}
and 
\begin{equation}
\label{thetaK2}
\theta_K + \i \epsilon_K \approx\frac{2\pi t_{12}\sigma_{xy}}{c \bar{n}(1+\alpha)(r_{12}-\alpha)} \:,
\end{equation}
where $\alpha = 2\pi\sigma/(c\bar{n})$. Note that Eq.~\eqref{thetaK2} is not valid when the difference $r_{12}-\alpha$ is close to zero, since in this case the condition $|r_+ - r_-| \ll |r_+ + r_-|$ does not hold. When, in addition to a small ratio $|\sigma_{xy}/\sigma|$, the parameter $\alpha$ is also small, i.e. $|\alpha| \ll 1$ and $|\alpha| \ll |r_{12}|$, it follows from Eqs.~\eqref{thetaF2} and \eqref{thetaK2}, that  the ratio of the Faraday and Kerr angles is constant, $\theta_K/\theta_F = t_{12}/r_{12}$. On the other hand, in the absence of dielectric contrast, when $n_1 = n_2 = \bar{n}$, and $r_{12} = 0$, $t_{12} = 1$, the frequency dependences of  the Faraday and Kerr angles differ, i.e. $\theta_F \approx 2\pi \mathrm{Re}\{ \sigma_{xy}\}/(c \bar{n})$, while  $\theta_K \approx -  2\pi\mathrm{Re}\{ \sigma_{xy}/\alpha \}/(c \bar{n}) $.

In a typical pump-probe experiment, see e.g. Ref.~[\onlinecite{Zhukov2007}], one measures the Faraday and Kerr rotation signals equal to the difference between the intensities of the transmitted and reflected beams, such as $I_{\omega, x'}^{(t)} - I_{\omega, y'}^{(t)}$ and $I_{\omega, \sigma+}^{(t)} - I_{\omega, \sigma-}^{(t)}$. Here, $(x', y')$ are the axes rotated by $\pi/4$ with respect to the initial $(x,y)$ frame, and $\sigma_\pm$ denotes right- and left-hand circular polarization. These signals are related to the rotation angles and ellipticities as
\begin{equation}
\label{expFar}
I_{\omega, x'}^{(t)} - I_{\omega, y'}^{(t)} = 2\theta_F T I_\omega\:,~~~I_{\omega, \sigma+}^{(t)} - I_{\omega, \sigma-}^{(t)} = 2\epsilon_F T I_\omega\:,
\end{equation}
and
\begin{equation}
\label{expKerr}
I_{\omega, x'}^{(r)} - I_{\omega, y'}^{(r)} = 2\theta_K R I_\omega\:,~~~I_{\omega, \sigma+}^{(r)} - I_{\omega, \sigma-}^{(r)} = 2\epsilon_K R I_\omega\:,
\end{equation}
where
\begin{equation}
\label{RT}
T = \frac{n_2 |\bar{t}|^2}{n_1}\:,~~~R = |\bar{r}|^2\:,
\end{equation}
$\bar{t} = (t_+ + t_-)/2$, $\bar{r} = (r_+ + r_-)/2$, and $I_\omega$ is the intensity of the incident probe field. Note that the dielectric contrast $n_1 \neq n_2$ is crucial for the experimental observation of the Kerr rotation signal, since the reflection coefficient $R$ for the free-standing 2D layer is proportional to the parameter $|\alpha|^2$, see Eq.~\eqref{rt}, which might be small~\cite{Zhukov2007}.

\section{Pump-induced transverse conductivity} \label{model}

Now, we develop a microscopic theory of the transverse conductivity $\sigma_{xy} (\omega,\Omega)$ induced by the circularly polarized pump field. The kinetics of 2D electrons driven by the pump and probe electric fields is described by the Boltzmann equation for the electron distribution function $f(\bm p, t)$
\begin{equation}
\label{kinetic}
\pderiv{f}{t} + e\left[\bm E_\Omega(t) + \bm E_\omega(t) \right] \cdot \pderiv{f}{\bm p} = \mathrm{St}f\:.
\end{equation}
Here, $\bm p$ is the electron momentum, $e$ is the electron charge and $\mathrm{St}f$ is the collision integral. The fields $\bm E_\Omega(t)$ and $\bm E_\omega(t)$ in Eq.~\eqref{kinetic} are electric fields experienced by the 2DEG, i.e. the sum of the incident and reflected fields at $z = 0$.
Equation~\eqref{kinetic} is valid in the classical regime, when $\hbar \omega$ and $\hbar \Omega$ are much less than the mean electron energy. 
We solve Eq.~\eqref{kinetic} by expanding the distribution function $f(\bm p, t)$ in the series in the electric field amplitude as follows:
\begin{multline}
f(\bm p, t) = f_0 + \left[ f_{1\omega} (\bm p) \e^{-\i\omega t} + f_{1\Omega} (\bm p) \e^{-\i\Omega t} + \mathrm{c.c.}\right] \\ + f_2 (\bm p)
 +  \left[ f_{2,\omega+\Omega} (\bm p) \e^{-\i(\omega+\Omega) t} + f_{2,\omega-\Omega} (\bm p) \e^{-\i(\omega-\Omega) t}  +\mathrm{c.c.} \right] 
 \\ + \left[ f_{3,\omega} (\bm p) \e^{-\i\omega t} + \mathrm{c.c.} \right]\:.
\end{multline}
Here, $f_0$ is the equilibrium distribution function, whereas the first-order corrections $f_{1\omega} \propto E_\omega$ and $f_{1\Omega} \propto E_\Omega$ determine Drude conductivity, responsible for ac electric currents oscillating at frequencies $\omega$ and $\Omega$, respectively. The second-order corrections are $f_2 \propto E_\Omega E^*_\Omega$, $f_{2,\omega+\Omega} \propto E_\omega E_\Omega$ and $f_{2,\omega-\Omega} \propto E_\omega E_\Omega^*$. The desired transverse current oscillating at $\omega$ is determined by the third-order correction $f_{3,\omega} \propto E_\omega E_\Omega E_\Omega^*$. 

Considering the term $e\left[\bm E_\Omega(t) + \bm E_\omega(t) \right] \cdot \partial f/\partial \bm p$ in Eq.~\eqref{kinetic} as a perturbation we obtain the following equations for the corrections to the distribution function:
\begin{subequations}
\label{kinetic2}
\begin{equation}
-\i\omega f_{1\omega} + e \bm E_\omega \cdot \pderiv{f_0}{\bm p} = \mathrm{St}~f_{1\omega}\:,  \label{f1w}
\end{equation}
\begin{equation}
 e \left( \bm E_{\Omega} \cdot \pderiv{f_{1\Omega}^*}{\bm p}
 + \bm E_{\Omega}^* \cdot \pderiv{f_{1\Omega}}{\bm p} \right)  = \mathrm{St}~f_2\:,  \label{f2} 
 \end{equation}
 \begin{multline}
\label{f2wplusW}
-\i(\omega + \Omega) f_{2, \omega + \Omega} 
+ e \left( \bm E_\omega \cdot \pderiv{f_{1\Omega}}{\bm p} +  \bm E_{\Omega} \cdot \pderiv{f_{1\omega}}{\bm p} \right)  
\\ = \mathrm{St}~f_{2,\omega + \Omega}\:, 
\end{multline}
\begin{multline}
\label{f3w}
-\i \omega f_{3,\omega} + e \bm E_\omega \cdot \pderiv{f_2}{\bm p} 
 +e \bm E_\Omega \cdot \pderiv{f_{2,\omega - \Omega}}{\bm p} 
\\ + e \bm E_\Omega^* \cdot \pderiv{f_{2, \omega + \Omega}}{\bm p}  
= \mathrm{St}~f_{3,\omega}\:. 
 \end{multline}
\end{subequations}
Equation for $f_{1\Omega}$ is obtained from Eq.~\eqref{f1w} by replacing $\omega$ with $\Omega$, and equation for $f_{2,\omega-\Omega}$ is obtained from Eq.~\eqref{f2wplusW} by replacing $\Omega$ with $-\Omega$ and making use of the relations $\bm E_{-\Omega} = \bm E_{\Omega}^*$, $f_{1,-\Omega} = f_{1\Omega}^*$.

In order to derive the $\sigma_{yx}$ component of the conductivity tensor, we calculate the transverse electric current $j_{\omega, y} = \sigma_{yx} E_{\omega, x}$ driven by the $x$-component of the probe field. The current reads
\begin{equation}
j_{\omega,y} = e \nu \sum_{\bm p} v_y f_{3,\omega}\:,
\end{equation}
where $\nu$ is the factor of spin and valley degeneracy. Multiplying Eq.~\eqref{f3w} by $v_y$ and averaging the result over the directions of $\bm p$, we obtain
\begin{multline}
\label{aver}
\aver{v_y f_{3,\omega}} = -e\tau_{1\omega} \aver{v_y \left( \bm E_\omega \cdot \pderiv{f_2}{\bm p} 
 + \bm E_\Omega \cdot \pderiv{f_{2,\omega - \Omega}}{\bm p} \right)}  \\
 -e\tau_{1\omega}  \aver{ v_y \bm E_\Omega^* \cdot \pderiv{f_{2, \omega + \Omega}}{\bm p}}\;,
\end{multline}
where $\aver{\dots}$ denotes averaging over the directions of $\bm p$, $\tau_{1\omega} = \tau_1/(1-\i\omega\tau_1)$, and $\tau_1^{-1} = -\aver{\bm v \mathrm{St} f}/\aver{\bm v f}$ is the energy-dependent momentum relaxation rate.
Summation of Eq.~\eqref{aver} over $\bm p$ and integration by parts yield
\begin{equation}
\label{jyw0}
j_{\omega, y} = e^2 \nu \sum_{\bm p} \left( f_2 \bm E_\omega + f_{2,\omega-\Omega} \bm E_\Omega +  f_{2,\omega+\Omega} \bm E_\Omega^* \right) \cdot \pderiv{(v_y \tau_{1\omega})}{\bm p}\:.
\end{equation}

We start with calculating $j_{\omega, y}$ for parabolic energy dispersion of electrons $\eps(\bm p) = |\bm p|^2/2m$. This dispersion is typical for low-energy electrons in III-V quantum wells, bilayer graphene, monolayers of transition metal dichalcogenides, etc. Calculating derivative in the right-hand side of Eq.~\eqref{jyw0}, one obtains
\begin{multline}
\label{jyw1}
j_{\omega, y} = e^2 \nu E_{\omega,x} \sum_{\bm p} v_x v_y \tau_{1\omega}' f_2 
\\ + \frac{e^2 \nu}{m} \sum_{\bm p}  (\eps \tau_{1\omega})' \left( f_{2,\omega-\Omega} E_{\Omega,y} +  f_{2,\omega+\Omega} E_{\Omega,y}^* \right) \\
+ \frac{e^2 \nu}{2} \sum_{\bm p}   \tau_{1\omega}' \left[ f_{2,\omega-\Omega}  \left( 2 v_x v_yE_{\Omega,x} - (v_x^2 - v_y^2) E_{\Omega,y}  \right) \right.
\\ \left. +  f_{2,\omega+\Omega} \left( 2 v_x v_y E_{\Omega,x}^* - (v_x^2 - v_y^2) E_{\Omega,y}^* \right)  \right]\:.
\end{multline}
Here, $(\dots)'$ denotes derivative over energy, and we took into account that $\bm E_\omega \parallel x$. The nature of the contributions to the ac current Eq.~\eqref{jyw1} is similar to the one discussed in Ref.~\cite{Durnev2021} for a static current. The first and the third contributions, proportional to $v_x v_y f_2$, $v_x v_y f_{2, \omega \pm \Omega}$ and  $(v_x^2 - v_y^2) f_{2, \omega \pm \Omega}$, are related to the optical alignment of electron momenta by the oscillating electric field. The second term, proportional to $(\eps \tau_{1\omega})'$, is related to the dynamic heating and cooling of 2DEG by the oscillating fields.

The first-order corrections to the distribution function are found from Eq.~\eqref{f1w} and read
\begin{equation}
\label{f1w_f1W}
f_{1\omega} = -e\tau_{1\omega} (\bm E_\omega \cdot \bm v) f_0'\:,~~~f_{1\Omega} = -e\tau_{1\Omega} (\bm E_\Omega \cdot \bm v) f_0'\:,
\end{equation}
where $\tau_{1\Omega} = \tau_1/(1-\i\Omega \tau_1)$. Calculation shows that the first term in Eq.~\eqref{jyw1} proportional to the time-independent correction $f_2$ vanishes for circularly polarized pump. Therefore, we do not consider this term in the following. Other second-order corrections are found by solving Eq.~\eqref{f2wplusW} with $f_{1\omega}$ and $f_{1\Omega}$ given by Eq.~\eqref{f1w_f1W}, which yields
\begin{multline}
\label{f2wWsol}
f_{2, \omega+\Omega} = \aver{f_{2, \omega+\Omega} } +  \frac12 e^2 E_{\omega,x}   \tau_{2,\omega + \Omega}  \left[ (\tau_{1\Omega}+\tau_{1\omega}) f_0' \right]'  \\ 
\times \left[ (v_x^2 - v_y^2)  E_{\Omega,x}  + 2 v_x v_y E_{\Omega,y} \right] \:.
\end{multline}
Here, 
$\aver{f_{2, \omega+\Omega}}$ is the zeroth angular harmonic of $f_{2, \omega+\Omega}$, 
$\tau_2^{-1} = -\aver{v_xv_y\mathrm{St}f}/\aver{v_xv_yf}$ is the energy-dependent relaxation rate of the second angular harmonic of the distribution function, and $\tau_{2,\omega+\Omega} = \tau_2/[1-\i(\omega+\Omega)\tau_2]$. 
 
We describe the relaxation of the zeroth angular harmonic of the distribution function $\aver{f(\bm p, t)}$ by the collision integral $\mathrm{St} \aver{f} = -(\aver{f} - f_0)/\tau_0$, where $\tau_0$ is the energy-independent relaxation time determined by the electron-electron scattering and energy-relaxation processes (e.g., caused by phonon scattering).
Equation~\eqref{f2wplusW} yields
\begin{equation}
\label{f2wplusW_aver}
\aver{f_{2, \omega+\Omega}} =  \frac{e^2 \tau_{0, \omega+\Omega}}{m} \left[\eps (\tau_{1\Omega}+\tau_{1\omega}) f_0' \right]'  E_{\omega,x} E_{\Omega,x}\:,
\end{equation}
where $\tau_{0, \omega + \Omega} = \tau_0/[1 - \i(\omega + \Omega) \tau_0]$.  The $f_{2,\omega - \Omega}$ function is found from $f_{2,\omega + \Omega}$ by replacing $\Omega$ with $-\Omega$ and using the relations $\tau_{1,-\Omega} = \tau_{1\Omega}^*$ and $\bm E_{-\Omega} = \bm E_{\Omega}^*$.

Finally, substituting Eqs.~\eqref{f2wWsol} and \eqref{f2wplusW_aver} into Eq.~\eqref{jyw1} for the current and calculating the sums, we obtain the transverse conductivity of the degenerate electron gas induced by the circularly polarized pump
\begin{equation}
\label{sigmaxy}
\sigma_{xy} (\omega, \Omega) = F(\omega, \Omega) - F(\omega, -\Omega)\:,
\end{equation}
where for parabolic spectrum
\begin{gather}
\label{par}
F^{(\rm par)}(\omega, \Omega) = - \frac{\i \sigma e^2 |\bm E_\Omega|^2 P_{\rm circ} [2 - \i(\omega+\Omega)\tau_1]}{2 m (1-\i\Omega\tau_1)}  \\
\times  \left[ (\eps_F \tau_{1\omega}'' + 2 \tau_{1\omega}')   \tau_{0, \omega+\Omega} 
- \eps_F (\tau_{1\omega}' \tau_{2,\omega + \Omega} )' - 2 \tau_{1\omega}' \tau_{2,\omega + \Omega}  \right] \nonumber \:.
\end{gather}
Here,  the relaxation times and its energy derivatives are taken at the Fermi energy $\eps_F$, $\sigma = e^2 n_e \tau_{1\omega}/m$ is the high-frequency conductivity,  and $n_e = \nu m \eps_F/(2\pi \hbar^2)$ is the electron density.

Similar calculations can be applied to 2DEG with linear energy dispersion, e.g. in graphene or HgTe/CdHgTe quantum wells of the critical thickness. Using $\eps(\bm p) = v_0 |\bm p|$ and performing calculations shown in App.~\ref{app}, one obtains $\sigma_{xy}$ given by Eq.~\eqref{sigmaxy} with 
\begin{gather}
F^{(\rm lin)}(\omega, \Omega) = -\frac{\i \sigma e^2 v_0^2  |\bm E_\Omega|^2 P_{\rm circ} [2 - \i(\omega+\Omega)\tau_1]}{4 \eps_F (1-\i\Omega\tau_1)} \nonumber
\\ \times \left[ \left( \eps_F \tau_{1\omega}'' + \tau_{1\omega}' - \frac{\tau_{1\omega}}{\eps_F}  \right) \tau_{0,\omega+\Omega}  - \eps_F(\tau_{1\omega}' \tau_{2,\omega +\Omega})' \right. \nonumber
\\ \left. - \tau_{1\omega}'\tau_{2,\omega +\Omega} + \tau_{1\omega} \left( \tau_{2,\omega+\Omega}' + \frac{ \tau_{2,\omega+\Omega}}{\eps_F} \right)
  \right]\:. 
\label{lin}
\end{gather}
Here, the high-frequency conductivity and the electron density are given by $\sigma= e^2 v_0^2 n_e \tau_{1\omega}/\eps_F$ and $n_e = \nu \eps_F^2/(4\pi \hbar^2 v_0^2)$.

Note that at $\omega = 0$, Eqs.~(\ref{sigmaxy} -- \ref{lin}) describe the static transverse photoconductivity of 2DEG and agree with the second line of Eq.~(16) in Ref.~[\onlinecite{Durnev2021}]. Conductivity given by Eqs.~\eqref{par} and \eqref{lin} 
is proportional to $| \bm E_\Omega|^2$, which is the square of the pump field at $z = 0$. $| \bm E_\Omega|^2$ is related to the intensity of the incident pump $I_\Omega = c n_1 [E_\Omega^{(i)}]^2/2\pi$ as $|\bm E_\Omega|^2 = 2\pi T(\Omega) I_\Omega/(c n_2)$, where $T$ is given by Eq.~\eqref{RT}.

\section{Discussion}

Equations~\eqref{thetaF2}, \eqref{thetaK2} and (\ref{sigmaxy} -- \ref{lin}) can be applied to calculate the photoinduced Faraday and Kerr rotation and ellipticity in different 2D systems, such as quantum wells, monolayer and bilayer graphene, transition metal dichalcogenide monolayers and other doped 2D materials. In this section we present results for two illustrative examples with linear and parabolic energy dispersion, monolayer and bilayer graphene, respectively. We also analyze the role of the dielectric contrast $(n_{2} - n_{1})/\bar{n}$ between the two dielectric media surrounding 2DEG on the rotation angles and ellipticities.

\subsection{2D layer on a substrate}

First, we consider the case of the 2D layer lying on a substrate by setting the refractive indices $n_1 = 1$ and $n_2 = 3$. In the discussion below Eq.~\eqref{thetaK2}, we showed that in case of a large dielectric contrast, the Kerr angle and ellipticity are related to the corresponding Faraday quantities as $\theta_K/\theta_F \approx t_{12}/r_{12}$, and $\epsilon_K/\epsilon_F \approx t_{12}/r_{12}$. Hence, for the chosen $n_1$ and $n_2$ we have $\theta_K \approx - \theta_F$ and $\epsilon_K \approx - \epsilon_F$, and in this subsection we discuss the Faraday angle and ellipticity only~\footnote{We assume that $n_{1,2}$ are frequency-independent in the considered frequency range.}.

\subsubsection{Parabolic spectrum. Bilayer graphene.}

\begin{figure}[htpb]
\includegraphics[width=0.47\textwidth]{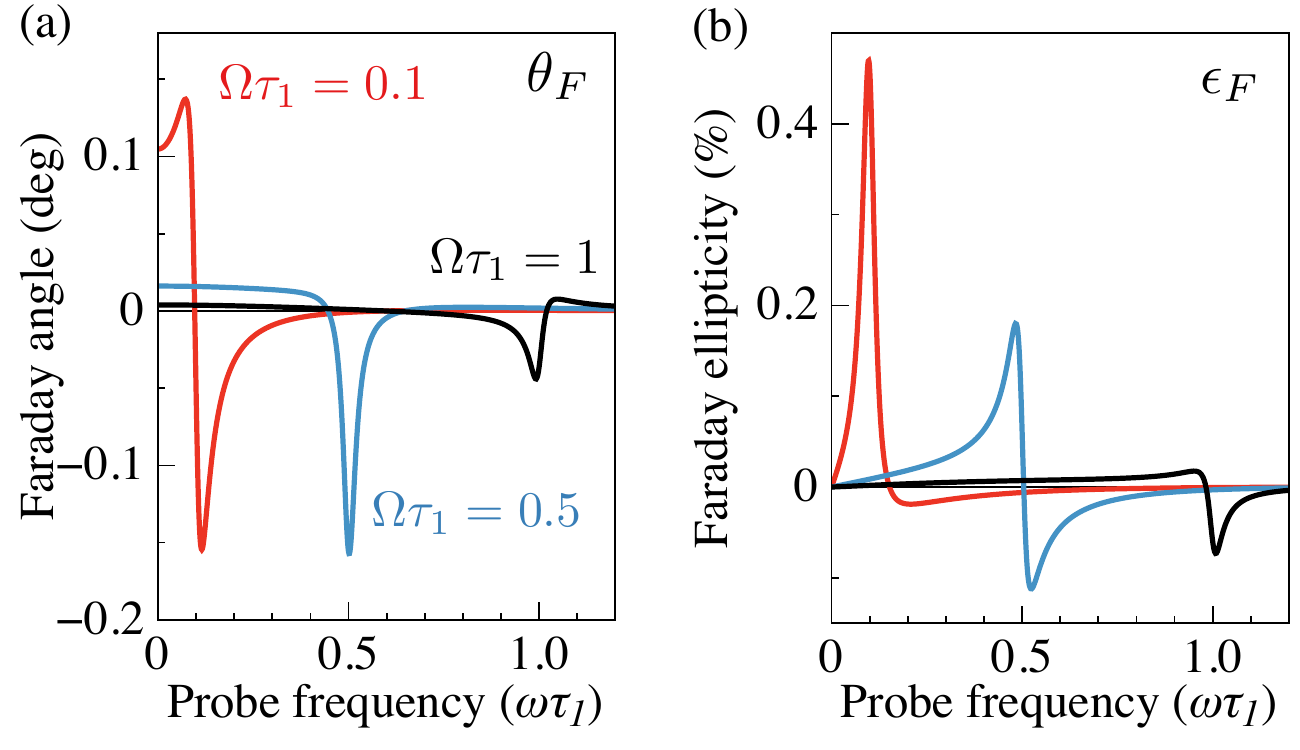}
\caption{\label{fig2} 
(a) Photoinduced Faraday rotation angle $\theta_F$ and (b) the accompanying ellipticity $\epsilon_F$ of the two-dimensional electron gas with \textit{parabolic} spectrum for a large dielectric contrast between the surrounding media. Three curves correspond to three values of the pump frequency: $\Omega \tau_1 = 0.1,~0.5,~1$. Sharp resonances at $\omega \approx \Omega$ occur. The curves are calculated after Eqs.~\eqref{thetaF2}, \eqref{sigmaxy} and \eqref{par} for the following parameters: $\tau_1 (\eps_F) = 0.1$~ps, $n_e = 10^{12}$~cm$^{-2}$, $\tau_0 = 5$~ps, $m = 0.03m_0$, $\tau_1 = 2\tau_2 \propto \eps$ (Coulomb scatterers), $I_\Omega = 1$~kW/cm$^2$, $n_1 = 1$, $n_2 = 3$ and $P_{\rm circ} = 1$. 
}
\end{figure}

Figure~\ref{fig2} shows the dependence of the calculated Faraday angle and the accompanying ellipticity for parabolic energy dispersion and a set of parameters relevant to bilayer graphene~\cite{Candussio2020}.  
It follows from Eq.~\eqref{par} that in case of the energy independent relaxation times $\tau_1$ and $\tau_2$, relevant for short-range scatterers, the transverse conductivity $\sigma_{xy}$ vanishes. Hence, the curves in Fig.~\eqref{fig2} are plotted for   unscreened Coulomb scatterers corresponding to $\tau_1 = 2\tau_2 \propto \eps$. We use the electron density $n_e = 10^{12}$~cm$^{-2}$ and momentum relaxation time $\tau_1(\eps_F) = 0.1$~ps, which results in $\eps_F \approx 39$~meV and $2\pi\sigma_0/(c\bar{n}) \approx 0.088$, where $\sigma_0 = e^2 n_e \tau_1/m$ is the static 2DEG conductivity. In the studied frequency range the transmission and reflection coefficients~\eqref{RT} lie in the range $T = 0.63 - 0.7$ and $R = 0.27 - 0.29$, respectively.

 The dependence of rotation angles and ellipticities on the probe frequency experiences sharp resonances in the region, where the probe frequency $\omega$ is close to the pump frequency $\Omega$. 
 At $\Omega\tau_1 \lesssim 1$ and pump intensity $I_\Omega = 1$~kW/cm$^2$, the Faraday angle at the resonance is $\theta_F \sim 0.1^\circ$, and the corresponding ellipticity $\epsilon_F \sim 0.1~\%$, see Fig.~\ref{fig2}. 
 Note that for such intensity, the inequality $|\sigma_{xy}| \ll |\sigma_{xx}|$ still holds so that we are still in the perturbative regime.
To study the shape of the resonances in more detail, we analyze the pump-induced conductivity, Eqs.~(\ref{sigmaxy} -- \ref{par}), at $\tau_0 \gg \tau_1$ relevant for 2DEG at low temperature, and $\Omega \tau_0 \gg 1$. In this case we have a sharp resonance in the conductivity, which shape for Coulomb scatterers is given by
\begin{equation}
\label{res_par}
\sigma_{xy} (\omega) \approx \frac{2 \i \sigma_0 e^2  \tau_1 \tau_{0} | \bm E_\Omega|^2 P_{\rm circ}}{m \eps_F[1 - \i(\omega-\Omega)\tau_0](1 + \Omega^2\tau_1^2)(1 - \i\Omega \tau_1)^3}  \:.
\end{equation}

Equation~\eqref{res_par} allows one to calculate the frequency dependence of the Faraday angle near the resonance. Substituting Eq.~\eqref{res_par} to Eq.~\eqref{thetaF2}, one obtains
\begin{multline}
\label{thetaF_res_par}
\theta_F (\omega) \approx \frac{4\pi \sigma_0}{c \bar{n}} \frac{e^2  \tau_1 \tau_0  | \bm E_\Omega|^2 P_{\rm circ}  }{m \eps_F} \\ \times \frac{\Omega\tau_1 (\Omega^2 \tau_1^2 - 3) + (\omega-\Omega) \tau_0 (3 \Omega^2 \tau_1^2 - 1)}{(1+\Omega^2\tau_1^2)^4[1 + (\omega-\Omega)^2 \tau_0^2]}\:.
\end{multline}
It follows from Eq.~\eqref{thetaF_res_par}, that depending on $\Omega \tau_1$, the resonance shape varies between Lorentzian and Lorentzian multiplied by $(\omega - \Omega)$, see Fig.~\ref{fig2}a.
Interestingly, the resonance width is given by the relaxation rate of the zeroth angular harmonic $\tau_0^{-1}$ rather than the momentum relaxation rate.  The magnitude of the resonance is determined by the product of $4\pi\sigma_0/(c\bar{n})$ and the dimensionless parameter $e^2  |\bm E_\Omega|^2   \tau_1 \tau_0/(m \eps_F)$ proportional to the intensity of the pump radiation. 

We note, that strictly at resonance, when $\omega = \Omega$, the developed theory is not applicable. In this case, one should consider a third-order response to the monochromatic electric field, since the pump and probe fields cannot longer be distinguished as in Eq.~\eqref{phen}. This situation corresponds to the self-induced rotation of electric field, when the field modifies dielectric properties of the 2D layer and, at the same time, experience rotation due to this modification. Such a self-induced rotation has been considered for graphene within a simplified relaxation model in Ref.~\cite{Glazov2014}. In App.~\ref{appB}, we calculate the third-order photocurrent induced by a monochromatic electric field being a sum of large circularly polarized and small linearly polarized contributions, see Eq.~\eqref{jy_res_final}.

\subsubsection{Linear spectrum. Single-layer graphene.}

\begin{figure}[htpb]
\includegraphics[width=0.49\textwidth]{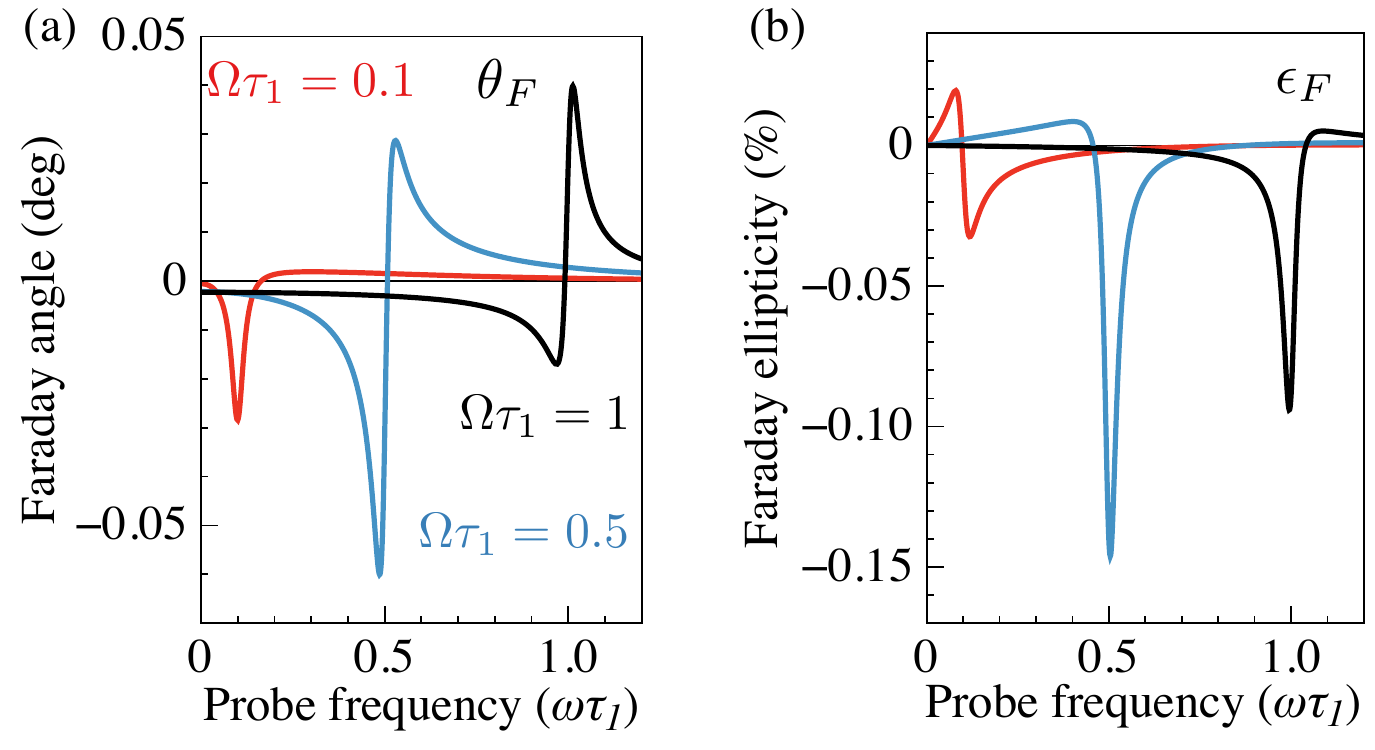}
\caption{\label{fig3} 
(a) Photoinduced Faraday rotation angle $\theta_F$ and (b) accompanying ellipticity $\epsilon_F$ of the two-dimensional electron gas with \textit{linear} spectrum for a large dielectric contrast $(n_2-n_1)/\bar{n}$ between the surrounding media. Three curves correspond to three values of the pump frequency: $\Omega \tau_1 = 0.1,~0.5,~1$. Sharp resonances at $\omega \approx \Omega$ occur. The curves are calculated after Eqs.~\eqref{thetaF2}, \eqref{sigmaxy} and \eqref{lin} for the following parameters: $\tau_1 (\eps_F) = 0.1$~ps, $n_e = 3\times10^{11}$~cm$^{-2}$, $\tau_0 = 5$~ps, $v_0 = 10^8$~cm/s, $\tau_1 = 2\tau_2 \propto \eps^{-1}$ (short-range scatterers), $I_\Omega = 1$~kW/cm$^2$, $n_1 = 1$, $n_2 = 3$ and $P_{\rm circ} = 1$. 
}
\end{figure}

Figure~\ref{fig3} shows the dependence of the calculated Faraday angle and the accompanying ellipticity for linear energy dispersion and a set of parameters relevant to monolayer graphene~\cite{Shimano2013}. For linear energy dispersion, the relaxation times are $\tau_1 = 2\tau_2 \propto \eps^{-1}$ for short-range scatterers and $\tau_1 = 3\tau_2 \propto \eps$ for Coulomb scatterers~\cite{Glazov2014}. It follows from Eq.~\eqref{lin} that both types of scatterers contribute to the transverse conductivity. For the calculations we use $n_e = 3\times10^{11}$~cm$^{-2}$ and  $\tau_1(\eps_F) = 0.1$~ps, which results in $\eps_F \approx 64$~meV and $2\pi\sigma_0/(c\bar{n}) \approx 0.071$. In that case, the transmission and reflection coefficients of the probe beam lie in the range $T = 0.65 - 0.71$ and $R = 0.26 - 0.28$, respectively.

As in the case of a bilayer graphene, the rotation angles and ellipticities in a single-layer graphene experience sharp resonances at $\omega \approx \Omega$.
The photoconductivity $\sigma_{xy}$ in the vicinity of resonance has the form
\begin{equation}
\label{res_lin}
\sigma_{xy}(\omega) \approx -\frac{\sigma_0 e^2 v_0^2   (3 - \i\Omega \tau_1) \Omega  \tau_1^2 \tau_{0} | \bm E_\Omega|^2 P_{\rm circ}}{2 \eps_F^2 [1 - \i(\omega-\Omega)\tau_0](1 + \Omega^2\tau_1^2)(1 - \i\Omega \tau_1)^3}  \:.
\end{equation}
Interestingly, Eq.~\eqref{res_lin} holds both for short-range and Coulomb scatterers.
Substituting Eq.~\eqref{res_lin} to Eq.~\eqref{thetaF2}, we obtain for the Faraday angle near the resonance:
\begin{multline}
\label{thetaF_res_lin}
\theta_F (\omega) \approx \frac{\pi \sigma_0}{c \bar{n}} \frac{e^2  v_0^2 \tau_1 \tau_0 | \bm E_\Omega|^2 P_{\rm circ} }{\eps_F^2} \\ \times \frac{\Omega\tau_1 [\Omega^4 \tau_1^4 + 6 \Omega^2 \tau_1^2 - 3  + 8\Omega\tau_1(\omega-\Omega)\tau_0]}{(1+\Omega^2\tau_1^2)^4[1 + (\omega-\Omega)^2 \tau_0^2]}\:.
\end{multline}
 The magnitude of the resonance is determined by the product of $\pi\sigma_0/(c\bar{n})$ and the dimensionless parameter $e^2  |\bm E_\Omega|^2   \tau_1 \tau_0/(m^* \eps_F)$ with the effective electron mass $m^* = \eps_F/v_0^2$ ($m^* \approx 0.01~m_0$ in our calculations).

\subsection{Free-standing monolayer graphene}

\begin{figure}[htpb]
\includegraphics[width=0.45\textwidth]{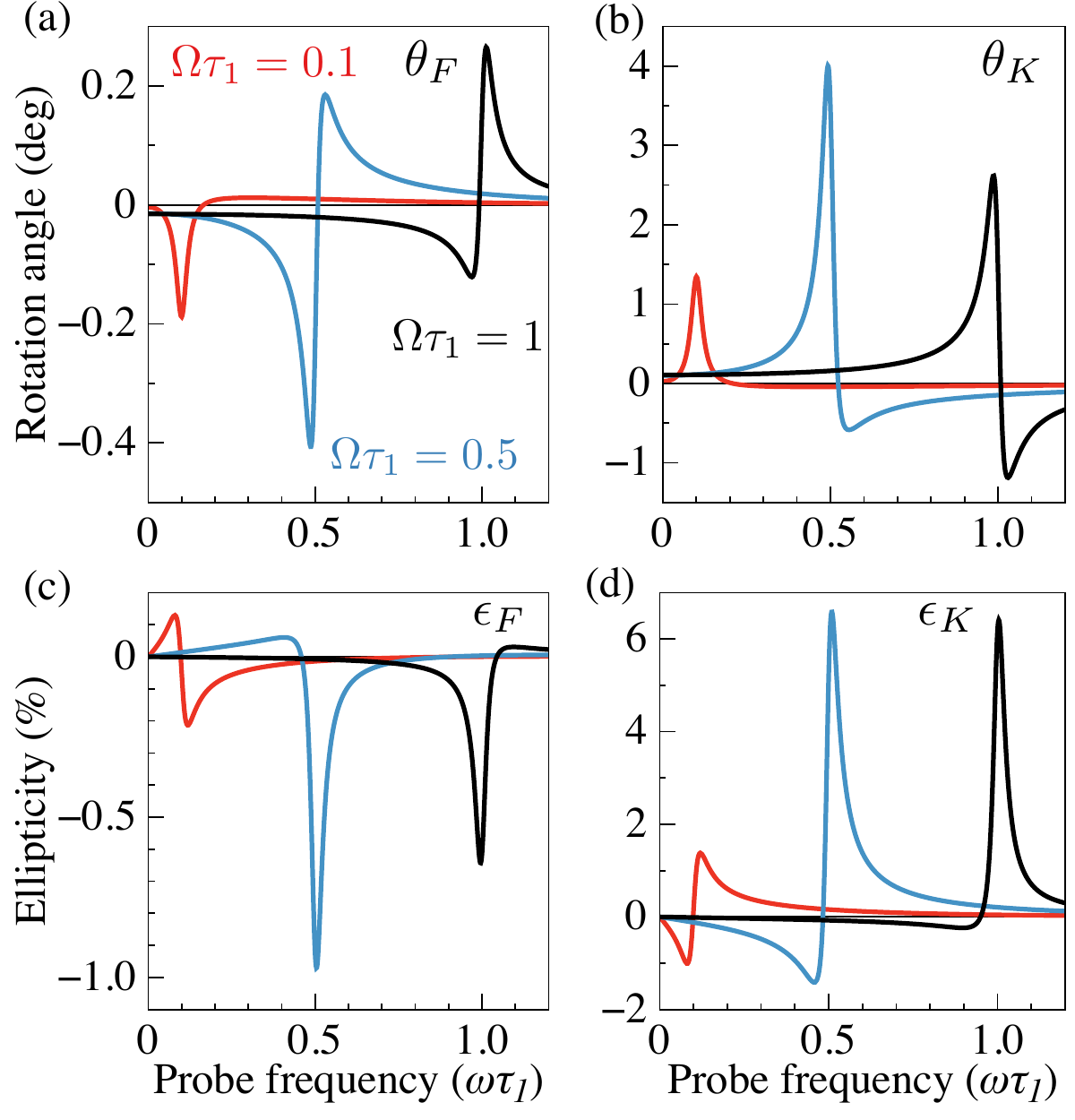}
\caption{\label{fig4} 
(a, b) Photoinduced Faraday and Kerr rotation angles $\theta_F$ and $\theta_K$  and (c, d) the accompanying ellipticities $\epsilon_F$ and $\epsilon_K$ of the two-dimensional electron gas in a \textit{free-standing} graphene. Three curves correspond to three values of the pump frequency: $\Omega \tau_1 = 0.1,~0.5,~1$. Sharp resonances at $\omega \approx \Omega$ occur. The curves are calculated after Eqs.~\eqref{thetaF2}, \eqref{thetaK2} and \eqref{lin} for the following parameters: $\tau_1 (\eps_F) = 0.1$~ps, $n_e = 3\times10^{11}$~cm$^{-2}$, $\tau_0 = 5$~ps, $v_0 = 10^8$~cm/s, $\tau_1 = 2\tau_2 \propto \eps^{-1}$ (short-range scatterers), $I_\Omega = 1$~kW/cm$^2$, $n_1 = n_2 = 1$ and $P_{\rm circ} = 1$. 
}
\end{figure}

In this section we consider a free-standing 2D layer by setting the refractive indices $n_1 = n_2 = 1$. 
In this case $r_{12} = 0$, $t_{12} = 1$, and as shown below Eq.~\eqref{thetaK2}, the Faraday and Kerr angles have different spectral dependences. Figure~\ref{fig4} shows the results of calculations for a free-standing monolayer graphene. The values of the rotation angles and ellipticities are larger for the free-standing layer than for the layer on a substrate, Figs.~\ref{fig2} and \ref{fig3}, for two reasons. First, the rotation angles and ellipticites are proportional to $1/\bar{n}$, see Eqs.~\eqref{thetaF2} and \eqref{thetaK2}. Second, the pump field at $z = 0$, $|\bm E_\Omega|^2 = 2\pi T(\Omega) I_\Omega/(c n_2)$, is larger at a given pump intensity. Moreover, the values of the Kerr angle and ellipticity are significantly larger than the corresponding Faraday values, since $\theta_F \propto \mathrm{Re}\{ \sigma_{xy}\}$, while $\theta_K \propto \mathrm{Re}\{ \sigma_{xy}/\alpha \}$ at $|\alpha| \ll 1$. Note that, however, the experimentally measured Kerr rotation signals, see Eq.~\eqref{expKerr}, are still small due to the small reflection from the free-standing layer.

The calculated Faraday rotation angles for graphene samples are $\sim 0.1^\circ - 1^\circ$ per 1~kW/cm$^2$ of the pump intensity, see Figs.~\ref{fig2}, \ref{fig3} and \ref{fig4}. Similar values of the Faraday angles were measured in monolayer and multilayer graphene in the terahertz and far-infrared frequency range at external magnetic field $B_z \sim 1$~T in Refs.~\cite{Crassee2011, Shimano2013}. The rotation angles can be further increased in high-mobility 2DEG in GaAs/AlGaAs quantum wells with larger values of $\tau_1$, see, e.g., Ref.~\cite{Suresh2020}.

\subsection{Synthetic magnetic field induced by pump}

The action of the circularly polarized pump on 2DEG can be described in terms of a synthetic magnetic field $B_{\rm syn}$. This field equals to an external magnetic field, which rotates the polarization plane by the same angle as the pump.
The Faraday angle in the presence of external magnetic field is given by Eq.~\eqref{thetaF2} with the Hall conductivity $\sigma_{xy} (B_z)$, which results in $\theta_F \sim (\omega_c \tau_1)2\pi\sigma_0/(c \bar{n})$, where $\omega_c = e B_z/mc$ is the cyclotron frequency.
By comparison with Eqs.~\eqref{thetaF_res_par} and \eqref{thetaF_res_lin} at $\Omega \tau_1 \sim 1$, one can estimate the synthetic magnetic field from $\omega_c \tau_1 \sim e^2  |\bm E_\Omega|^2   \tau_1 \tau_0/(m \eps_F)$, which yields
 \begin{equation}
 B_{\rm syn} \sim \frac{e c |\bm E_\Omega|^2 \tau_0}{\eps_F}\:.
 \end{equation}
 Note that the value of $B_{\rm syn}$ is quite universal, since it does not depend on the electron mobility and energy dispersion. It depends, however, on the energy relaxation time $\tau_0$ and, hence, should increase with decreasing temperature.
 
Synthetic magnetic field induced by the pump with intensity $I_\Omega = 1$~kW/cm$^2$ at $\eps_F = 50$~meV and $\tau_0 = 10$~ps is $B_{\rm syn} \sim 0.1$~T. This value increases with the growth of radiation intensity and may reach 1~T for several kW/cm$^2$ terahertz and far-infrared radiation, which is used for spectroscopy of electron gas in graphene~\cite{McIver2020, Candussio2022}. Note that $B_{\rm syn}$ is significantly (several orders of magnitude) larger than the actual magnetic field induced by the orbital currents being the source of the inverse-Faraday magnetization~\cite{Hertel2006, Candussio2020}.

\section{Summary}

To summarize, we have studied theoretically the pump-probe Faraday and Kerr rotation due to the orbital magnetization in the two-dimensional electron gas (2DEG).
We have shown that the circularly polarized electric field of the terahertz-range pump results in the transverse 
conductivity $\sigma_{xy} (\omega, \Omega)$ of 2DEG, which is proportional to the pump intensity and depends on both the probe and pump frequencies $\omega$ and $\Omega$, respectively. This pump-induced anisotropy of conductivity results in the circular birefringence and dichroism for a probe field.
We have derived analytical expressions for $\sigma_{xy} (\omega, \Omega)$ and the corresponding Faraday and Kerr rotation angles for parabolic and linear energy dispersion of 2D electrons and arbitrary scattering potential. We have shown that at $\omega \approx \Omega$ rotation angles are resonantly enhanced, reaching $0.1^\circ - 1^\circ$ for 1~kW/cm$^2$ of the pump intensity in graphene samples at $\Omega \tau_1 \sim 1$, where $\tau_1$ is the momentum relaxation time. Similar values of the Faraday angles were measured in monolayer and multilayer graphene in the terahertz and far-infrared frequency range in an external magnetic field $B_z \sim 1$~T~\cite{Crassee2011, Shimano2013}. 
The calculated Faraday and Kerr angles are governed by the momentum and energy relaxation of 2D electrons, and hence, can elucidate mechanisms and rates of electron relaxation processes in pump-probe experiments.

\acknowledgments
The author thanks S. A. Tarasenko and M. M. Glazov for fruitful discussions. The work was supported by the Russian Science Foundation (Project No. 21-72-00047).

\appendix
\section{Transverse photoconductivity of 2DEG with linear energy spectrum} \label{app}

Here, we calculate pump-induced transverse conductivity for electrons with linear energy dispersion $\eps = v_0 p$. We start with the general equation for the current~\eqref{jyw0}. Calculating derivative on the right-hand side of Eq.~\eqref{jyw0} one obtains
\begin{multline}
\label{jywlin}
j_{\omega, y} = e^2 \nu E_{\omega,x} \sum_{\bm p} v_x v_y \eps \left(\frac{\tau_{1\omega}}{\eps}\right)' f_2 \\
+ \frac{e^2 v_0^2 \nu}{2} \sum_{\bm p} \frac{(\eps \tau_{1\omega})'}{\eps}   \left( f_{2,\omega-\Omega} E_{\Omega,y} +  f_{2,\omega+\Omega} E_{\Omega,y}^* \right) \\
+ \frac{e^2 \nu}{2} \sum_{\bm p}  \eps \left( \frac{\tau_{1\omega}}{\eps} \right)' \left[ f_{2,\omega-\Omega}  \left( 2 v_x v_yE_{\Omega,x} - (v_x^2 - v_y^2) E_{\Omega,y}  \right)  \right.
\\ \left. +  f_{2,\omega+\Omega} \left( 2 v_x v_y E_{\Omega,x}^* - (v_x^2 - v_y^2) E_{\Omega,y}^* \right)  \right]\:.
\end{multline}
The first contribution in Eq.~\eqref{jywlin} proportional to $f_2$ vanishes for circularly polarized pump.
The first-order corrections to the distribution function coincide with the ones given by Eq.~\eqref{f1w_f1W}, whereas the second-order correction $f_{2, \omega+\Omega}$ has the form
\begin{multline}
\label{lin1}
f_{2, \omega+\Omega} = \aver{f_{2, \omega+\Omega}} +  \frac{e^2 E_{\omega,x} }{2}  \tau_{2,\omega +\Omega}  \eps \left[ \frac{(\tau_{1\Omega}+\tau_{1\omega}) f_0'}{\eps} \right]' \\
\times \left[ (v_x^2 - v_y^2)  E_{\Omega,x}  + 2 v_x v_y E_{\Omega,y} \right] \:,
\end{multline}
where
\begin{equation}
\label{lin2}
\aver{f_{2, \omega+\Omega}} = \frac{e^2 v_0^2 \tau_{0, \omega+\Omega}}{2\eps}  \left[ \eps ( \tau_{1\omega} + \tau_{1\Omega}) f_0' \right]'  E_{\omega,x} E_{\Omega,x}\:.
\end{equation}
The $f_{2, \omega-\Omega}$ function is obtained from Eqs.~\eqref{lin1}, \eqref{lin2} by replacing $\Omega$ with $-\Omega$ and using the relations $\tau_{1,-\Omega} = \tau_{1\Omega}^*$ and $\bm E_{-\Omega} = \bm E_{\Omega}^*$.
Finally, substituting $f_{2, \omega\pm\Omega}$ given by Eqs.~(\ref{lin1}--\ref{lin2}) into Eq.~\eqref{jywlin} for the current and calculating the sums we obtain Eqs.~\eqref{sigmaxy} and \eqref{lin} of the main text.

\section{Transverse photoconductivity at coinciding pump and probe frequencies} \label{appB}

In this section, we calculate third-order response similar to Eq.~\eqref{phen} but at coinciding pump and probe frequencies, $\omega = \Omega$. Electric field at the 2DEG plane $\bm E(t) = \bm E \e^{-\i\omega t} + \mathrm{c.c.}$ is a sum of large circularly polarized (pump) and small linearly polarized (probe) contributions:
\begin{equation}
\label{E_res}
E_{x} = \frac{E_1}{\sqrt{2}} + E_2\:,~~~E_{y} = \i P_{\rm circ} E_1/\sqrt{2}\:,
\end{equation}
where $P_{\rm circ} = \pm 1$ and $E_2 \ll E_1$.

We search the electron distribution function $f(\bm p, t)$ in the form
\begin{multline}
f(t) = f_0 + \left[ f_{1} (\bm p) \e^{-\i\omega t} + \mathrm{c.c.} \right] + f_2 (\bm p) \\
+  \left[ \tilde{f}_{2} (\bm p) \e^{-2\i\omega t} + \mathrm{c.c.} \right] + \left[ f_{3} (\bm p) \e^{-\i\omega t} + \mathrm{c.c.} \right]\:,
\end{multline}
where corrections to the distribution function satisfy the following equations
\begin{subequations}
\label{kinetic3}
\begin{align}
\label{F1}
-\i\omega f_{1} + e \bm E \cdot \pderiv{f_0}{\bm p} = \mathrm{St}~f_{1}\:,  \\
%
\label{F2}
 e \left( \bm E \cdot \pderiv{f_{1}^*}{\bm p} + \bm E^* \cdot \pderiv{f_{1}}{\bm p} \right) = \mathrm{St}~f_2\:,  \\
\label{tildeF2}
-2\i\omega \tilde{f}_{2} + e \bm E  \cdot \pderiv{f_{1}}{\bm p} = \mathrm{St}~\tilde{f}_2\:,  \\
\label{F3}
-\i \omega f_{3} + e \bm E \cdot \pderiv{f_2}{\bm p}  + e \bm E^* \cdot \pderiv{\tilde{f}_2}{\bm p}  = \mathrm{St}~f_{3}\:. 
\end{align}
\end{subequations}

The transverse electric current is determined by the third-order correction $f_3$ and reads
\begin{equation}
j_{\omega, y} = e \sum_{\bm p} v_y f_3 =  e^2 \sum_{\bm p} \left( f_2 \bm E +  \tilde{f}_2\bm E^* \right) \cdot \pderiv{(v_y \tau_{1\omega})}{\bm p}\:.
\end{equation}
Taking derivative in the right-hand side for the case of linear dispersion and simplifying, we obtain
\begin{multline}
\label{jy_res}
j_{\omega, y} = e^2  v_0^2 \sum_{\bm p} \left[ \frac{\tau_{1\omega}}{\eps}  + \frac{\eps}{2}\left(\frac{\tau_{1\omega}}{\eps} \right)' \right]  f_{2} E_{y} \\
+ \frac{e^2 }{2} \sum_{\bm p}  \eps \left( \frac{\tau_{1\omega}}{\eps} \right)'  f_{2} \left[  2 v_x v_y E_{x} - (v_x^2 - v_y^2) E_{y} \right] \\
 + e^2  v_0^2 \sum_{\bm p} \left[ \frac{\tau_{1\omega}}{\eps}  + \frac{\eps}{2}\left(\frac{\tau_{1\omega}}{\eps} \right)' \right]  \tilde{f}_{2} E_{y}^* \\
+ \frac{e^2 }{2} \sum_{\bm p}  \eps \left( \frac{\tau_{1\omega}}{\eps} \right)'  \tilde{f}_{2} \left[  2 v_x v_y E_{x}^* - (v_x^2 - v_y^2) E_{y}^* \right]\:.
\end{multline}

By solving Eqs.~\eqref{F2} and \eqref{tildeF2} with the use of Eq.~\eqref{f1w_f1W}, we obtain
\begin{multline}
\label{f2res}
f_{2} =  e^2  \tau_{2} \mathrm{Re} \left\{ \eps \left( \frac{\tau_{1\omega} f_0'}{\eps} \right)' \right\}  \left[ (v_x^2 - v_y^2)  S_1  + 2 v_x v_y S_2 \right] \\
+ e^2 v_0^2 \tau_{0} S_0 \mathrm{Re} \left\{ \frac{(\eps \tau_{1\omega} f_0')'}{\eps} \right\}\:, 
\end{multline}
and
\begin{multline}
\label{tildef2res}
\tilde{f}_{2} =  \frac{e^2  \tau_{2,2\omega}}{2} \eps \left( \frac{\tau_{1\omega} f_0'}{\eps} \right)'  \left[ (v_x^2 - v_y^2) s_1  + 2 v_x v_y s_2 \right] \\
+ \frac{e^2 v_0^2 \tau_{0,2\omega} s_0}{2\eps} (\eps \tau_{1\omega} f_0')'  \;.
\end{multline}
Here, $S_0 = |\bm E|^2$, $S_1 = |E_x|^2 - |E_y|^2$, $S_2 = E_x E_y^* + E_x^* E_y$ are the Stokes parameters, and $s_0 = E_x^2 + E_y^2$,  $s_1 = E_x^2 - E_y^2$, $s_2 = 2E_x E_y$. By substituting Eqs.~\eqref{f2res}, \eqref{tildef2res} and \eqref{E_res} in Eq.~\eqref{jy_res} for the current, performing summation over $\bm p$ and simplifying, we finally obtain
\begin{multline}
\label{jy_res_final}
j_{\omega, y} =  -\frac{\i \sigma e^2 v_0^2 P_{\rm circ} E_1^2 E_2}{\eps_F} \left\{ \left( \frac{2 \tau_0}{1+\i \omega \tau_1} - \tau_{0,2\omega} \right) A \right.
 \\
\left. - \frac{\tau_2 A + \eps_F\tau_2' B}{1+\i\omega \tau_1}   + \frac{ 3}{2} \left( \tau_{2,2\omega} A + \eps_F\tau_{2, 2\omega}' B \right) \right\} \:, 
\end{multline}
where 
\begin{equation}
A = \eps_F \tau_{1\omega}'' + \tau_{1\omega}' - \frac{\tau_{1\omega}}{\eps_F}\:,~~~B = \tau_{1\omega}' - \frac{\tau_{1\omega}}{\eps_F}\:.
\end{equation}
Here, we only left contributions to the current proportional to $E_1^2 E_2$.

Note that, for a simplified relaxation model with relaxation times $\tau_0 = \tau_1 = \tau_2$ and independent of energy, the current given by Eq.~\eqref{jy_res_final} coincides with Eq.~(69) of Ref.~\cite{Glazov2014}.

\bibliographystyle{apsrev4-1-customized}
\bibliography{bibliography}

\end{document}